\documentclass[preprint,authoryear,12pt]{elsarticle}

\usepackage{graphicx}

\usepackage{amssymb}
\newcommand{\apj}{    {\rm Astrophys. J.}\ }
\newcommand{\apjl}{   {\rm Astrophys. J. Lett.}\ }

\newcommand{\solphys}{{\rm Solar Phys.}\ }

\journal{JASTP, special issue}

\begin{document}

\begin{frontmatter}



\title{Solar Cycle Variations of Rotation and Asphericity in the Near-Surface Shear Layer}


\author{A.~G. Kosovichev\corref{cor1}\fnref{1}}
\ead{akosovichev@solar.stanford.edu}
\author{J.~P. Rozelot\fnref{2}}
\ead{jp.rozelot@orange.fr}

\address[1]{New Jersey Institute of Technology, Newark, NJ 07103, USA }
\address[2]{Nice University, OCA-CNRS, 77 Chemin des Basses Mouli\`eres, 06130 Grasse (F)}

\begin{abstract}
The precise shape of the Sun is sensitive to the influence of gravity, differential rotation, local turbulence and magnetic fields. So its precise measurement is a long-standing astrometric objective. It has been previously shown by different methods that the solar shape exhibits asphericity that evolves with the solar cycle. Thanks to the Michelson Doppler Imager (MDI) on Solar and Heliospheric Observatory (SoHO) and the Helioseismic and Magnetic Imager (HMI) aboard NASA's Solar Dynamics Observatory (SDO), and their capability to observe with an unprecedented accuracy  the surface gravity oscillation (f) modes, it is possible to extract information concerning the coefficients of rotational frequency splitting, $a_1$, $a_3$ and $a_5$, that measure the latitudinal differential rotation, together with the $a_2$, $a_4$ and $a_6$ asphericity coefficients. Analysis of these helioseismology data with time for almost two solar cycles, from 1996 to 2017, reveals a close correlation of the $a_1$ and $a_5$ coefficients with the solar activity, whilst $a_3$ exhibits a long-term trend and a weak correlation with the solar activity in the current solar cycle indicating a substantial change of the global solar rotation, potentially associated with a long-term evolution of the solar cycles. Looking in more details, the asphericity coefficients, $a_2$, $a_4$ and $a_6$ are more strongly associated with the solar cycle when applying a time lag of respectively 0.1, 1.6 and -1.6 years. The magnitude of $a_6$-coefficient varies in phase with the sunspot number (SN), but its amplitude is ahead of the SN variation. The latest measurements made in mid 2017 indicate that the magnitude of the $a_6$-coefficient has probably reached its minimum; therefore, the next solar minimum can be expected by the end of 2018 or in the beginning of 2019. The so-called ``seismic radius" in the range of f-mode angular degree: $\ell= 137-299$ exhibits a temporal variability in anti-phase with the solar activity; its relative value decreased by  $\sim 2.3\times 10^{-5}$ in Solar Cycle 23 and $1.7\times 10^{-5}$ in Cycle 24. Such results will be useful for better understanding the physical mechanisms which act inside the Sun, and so, better constrain dynamo models for forecasting the solar cycles.
\end{abstract}


\end{frontmatter}

\section{Introduction}\label{In}
Helioseismology is a powerful tool to study the structure and dynamics of the Sun's interior. Substantial progress has been made mainly due to the help of uninterrupted observations since 1995 by the GONG (Global Oscillations Network Group) and by the SoHO/MDI and SDO/HMI space instruments.   Global helioseismology provides important information about the physics of localized structures beneath the surface of the Sun, which has led to major advances in our understanding of the solar dynamics  such as detection of the sharp radial gradient of the differential rotation at the base of the convection zone (the tachocline) \citep{Kosovichev1996}, the near surface rotational shear layer (NSSL) \citep{Schou1998} that occupies approximately the top 5\% of the solar radius and presumably plays a fundamental role in the solar dynamo \citep{Godier2001,Brandenburg2005,Pipin2011}, as well as to measurements of the seismic solar radius and its variations \citep{Schou1997,Dziembowski1998,Dziembowski2001,Lefebvre2005,Lefebvre2007}. Global helioseismology continues to produce new and challenging results along with local helioseismology, another means to access the internal dynamics and properties of the Sun \citep[e.g.][]{Kosovichev2016}. Of particular interest are the temporal variations of the surface gravity oscillation (f) modes that probe the dynamics and structure of the NSSL, as well as the solar seismic radius. 

The surface gravity waves are excited by turbulent convection in the upper convective layer of the Sun. These fundamental modes, called f-modes, peak near frequency 3 mHz; and the envelope peak occurs near $\ell = 880$, covering the degree range of 100-1200 in the observed power spectrum. Our analysis is focused on the low-frequency medium-degree f-modes in the range of $\ell=137-299$, which were observed during the whole period of the MDI and HMI observations. In this range, the f-modes kinetic energy is concentrated within a layer of approximately 15~Mm deep, or about 2\% of the solar radius \citep[see Fig.2 of][]{Lefebvre2005}. The frequencies of these modes are affected by the surface magnetism and temperature/sound-speed changes not as significantly as the frequencies of acoustic (p) modes, are reflect large-scale variations in the near-surface shear layer. In addition, because these modes do not propagate to the surface they are characterized by narrow line widths in the oscillation power spectrum. This allows to measure the mode frequencies and splitting coefficients with high precision. Also, to reduce potential effects of strong surface variations caused by active regions we consider only the frequency splitting coefficient of low-degree, $a_1-a_6$, corresponding to latitudinal variations described by Legendre polynomials of degree $j=1-6$; the `odd' coefficients describe the differential rotation, and the `even' coefficients reflect the global asphericity. The f-mode asphericity coefficients are mostly due to Lagrangian variations of the radius of the subsurface layers, although there still might be some direct effect of magnetic field and turbulent stresses.  Up to now, the physics of the temporal dependence of the asphericity coefficients has not been fully analyzed and understood. Nevertheless, a correlation analysis presented in this paper shows interesting relationships and encourages further observational and theoretical studies using the great amount of available global helioseismology data. 

\section{Data and results}

The time-series of solar oscillations used here have been provided by the two space missions SOHO (Solar and Heliospheric Observatory) \citep{Scherrer1995} and SDO (Solar Dynamics Observatory) \citep{Scherrer2012}. The f-modes in the medium-degree range are not observed from the ground because of their low amplitude. 
The data from both instruments are available on-line from the SDO JSOC (Joint Science Operations Center) archive: http://jsoc.stanford.edu \citep{Larson2016}. The mode frequency analysis is performed using 72-day series of full-disk Dopplergrams. The HMI high-resolution data are specially prepared to match the spatial resolution of the MDI Medium-$\ell$ Structure Program \citep{Kosovichev1996}. We study here the whole time span ranging from April 30, 1996, to June 4, 2017. The total number of the frequency datasets combined from the two instruments for our analysis was 105.  

We recall the basic definition of f-mode frequency $\nu_{n,\ell,m}$, where $\ell$ and $m$ are the spherical harmonic degree and the azimuthal order respectively, and $n$ is the radial order which is zero for f-modes. The $a$-coefficients for each $l$ obtained from the SOHO-MDI and SDO-HMI data, are defined by
\begin{equation}
\nu_{n,\ell,m} = \nu_{n,\ell} + \sum^{6}_{j=1} a_{j}(n,\ell) {\cal P}^{\ell}_{j} (m),
\end{equation}
where $\nu_{n,\ell}$ is the mean (`central') multiplet frequency, and {\slshape $P^{\ell}_{j}(m)$ } are orthogonal
polynomials of degree $j$ defined by ${\cal P}_{j}^{\ell}(\ell) = \ell$ ,  and $\sum^{\ell}_{m=-\ell} {\cal P}^{\ell}_{i} (m) P^{l}_{j} (m) =0$ for $i\neq j $ \citep{Ritzwoller1991}.

The odd coefficients ($a_1$, $a_3$, $a_5$) reflect the rotational part of the fine structure in the spectrum of solar oscillations, whilst the even coefficients ($a_2$, $a_4$, $a_6$) reflect the solar asphericity \citep[e.g][]{Dziembowski2004}. The higher degree coefficients calculated up to degree 36 are used to study migrating zonal flows (`torsional oscillations') \citep{Kosovichev1997a}. In this work we use only the low-order splitting coefficients ($a_1-a_6$) and also the central mode frequencies $\nu_{n,\ell}$ that are calculated with higher precision and provided as a separate dataset.   

\begin {figure} []
\begin {center}
\includegraphics[width=\linewidth]{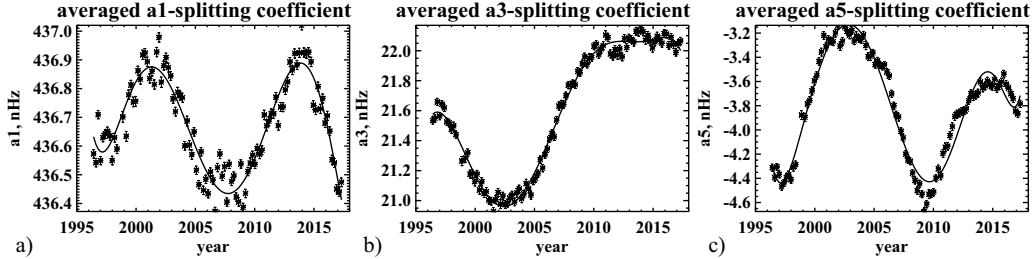}
\caption {Rotation coefficients $a_1$, $a_3$ and $a_5$ as deduced from the analysis of the f-modes extracted from the Michelson Doppler Imager (MDI) and Helioseismic and Magnetic Imager (HMI) aboard Solar and Heliospheric Observatory (SoHO) and Solar Dynamics Observatory (SDO) data from 1996 to 2017. The error bars show three standard deviations calculated from the formal error estimates.}
\label{rotation}
\end {center}
\end {figure}

\begin{figure}
	\centering
	\includegraphics[width=\linewidth]{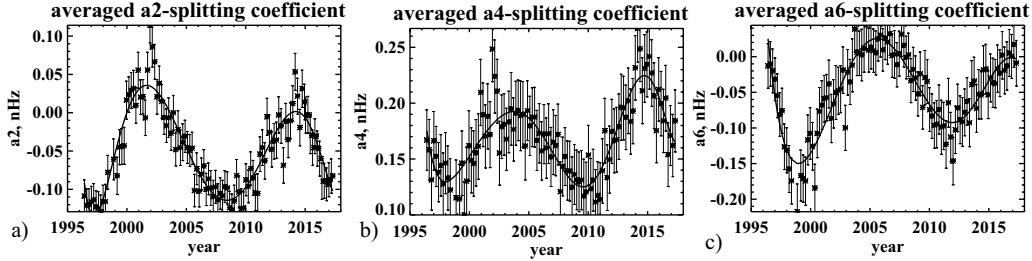}	
	\caption{Asphericity coefficients $a_2$, $a_4$ and $a_6$ as deduced from the analysis of the f-modes extracted from the MDI and HMI data from 1996 to 2017. The error bars show three standard deviations calculated from the formal error estimates.} 
	\label{asphericity}
\end{figure}

\begin{figure}[h]
	\centering
	\includegraphics[width=0.9\linewidth]{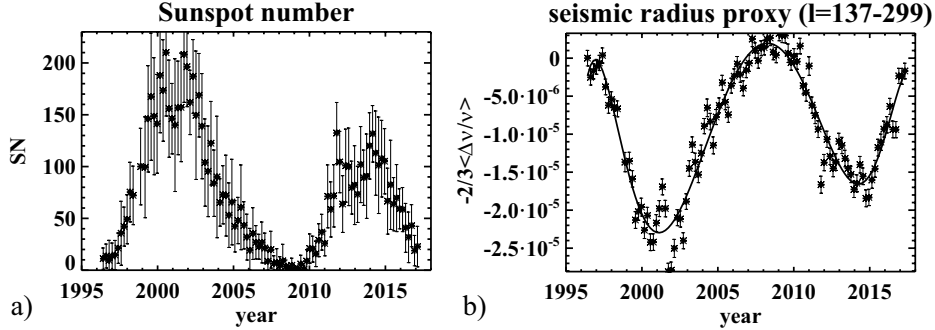}	

		\caption{a) The sunspot number, SN, averaged for the 72-day periods corresponding to the intervals of the helioseismic analysis. b) Variations of the seismic radius proxy (Eq.~\ref{eqradius}) relative to the first measurement in 1996, as deduced from the analysis of the f-modes extracted from the MDI and HMI data from 1996 to 2017. The relative amplitude modulation of about $-2.3\times 10^{-5}$ in Solar Cycle 23 and about $-1.7\times 10^{-5}$ in Cycle 24 is clearly in anti-phase with the solar activity. The error bars show three standard deviations calculated using observational error estimates of the mean f-mode frequencies.}
		\label{radius}
\end{figure}

To study the variations with the solar cycle we averaged the coefficients over a common subset of modes which were observed in all 72-day periods.  The total number of f-modes in the common subset which covers the range of $\ell$ from 137 to 299 is 152. The averaged odd coefficients are plotted in Figure~\ref{rotation}, and the averaged even coefficients are plotted in Figure~\ref{asphericity}. The error bars are calculated using the formal error estimates provided for the measurements of individual modes \citep{Larson2015}. The solid curves show eighth-degree polynomials fitted to the averaged coefficients.  For comparison with solar activity, we used the daily sunspot number data from the WDC-SILSO, Royal Observatory of Belgium, Brussels, to calculate the 72-day averages which correspond exactly to the helioseismology intervals. These averages are shown in Fig.~\ref{radius}$a$.  
   
All the coefficients show strong changes over the 21-year period that spans two solar activity cycles, but not all of them show corresponding cyclic variations. In particular, it is striking that while $a_1$ and $a_5$ show two peaks corresponding to the sunspot maxima, the $a_3$-coefficient shows a  negative peak corresponding to maximum of Solar Cycle 23, but there is no such peak for the current Cycle 24. From the helioseismology theory it follows the $a$-coefficients correspond to the differential rotation law expressed in terms of the associate Legendre functions, $P_n^m(\theta)$, \citep{Kosovichev1988}: 
\begin{equation}
\Omega(\theta)/2\pi=\left<a_1\right>-\left[\frac{2}{3}\left<a_3\right>P^1_3(\theta)+\frac{8}{15}\left<a_5\right>P^1_5(\theta)\right]/\sin\theta,
\end{equation}
The helioseismic odd $a$-coefficients can be compared with the corresponding coefficients of the surface rotation law, which were determined  by \citet{Snodgrass1984}: $a_1^s=435$~nHz, $a_3^s=-21$~nHz, and $a_5^s=-4$~nHz.

The observed variations of the coefficients (Fig.~\ref{rotation}) show that the magnitude of $a_1$ increased during the solar maximum, but the magnitude of $a_3$ and $a_5$ decreased, meaning that they change in anti-phase with the solar activity cycle.  This also means that the angular variation of the subsurface differential rotation was reduced (it became more solid-like). However, it is intriguing that during the Cycle 23 the $a_3$ coefficient significantly decreased but remains almost constant during the Cycle 24. Although the behavior of the last term ($a_5$ coefficient) in both cycles was qualitatively similar. This shows that there is a substantial change in the global dynamics of the Sun during the current relatively weak cycle. An unusual behavior of the migrating zonal flows, so-called ``torsional oscillations" has been noticed in this solar cycle from global and local helioseismology measurements \citep{Howe2013,Kosovichev2016}. 

The even $a$-coefficients shown in Fig.~\ref{asphericity} correspond to the solar asphericity in terms of the Legendre polynomials $P_2(\cos\theta)$, $P_4(\cos\theta)$, and $P_6(\cos\theta)$ \citep{Kuhn1988}. All of them show a two-peak structure corresponding to the two solar maxima of 2000-2002 and 2012-2015. However, it appears that the asphericity of the Sun dramatically changes from the solar minimum to maximum. During the solar minima the asphericity was dominated by the $P_2$ and $P_4$ terms, and $P_6$ contribution was negligible. But, during the activity maxima $P_2$ became small, but the values of $P_4$, and, in particular, $P_6$ substantially increased. This means that the $P_2$ asphericity changes in anti-phase with the solar cycle, while the $P_4$ and $P_6$ change in phase. At this point, we cannot make a physical interpretation of these variations, but, it is interesting that the primary terms describing the solar oblateness is reduced practically to zero, meaning that the asphericity associated with the solar activity during the maxima almost compensates the rotational distortion expressed in terms of the second-order Legendre polynomial.

To quantify the temporal behavior of the f-mode coefficients we calculated their cross-correlations with the sunspot number.
The cross correlations with variable time lag were calculated for the whole 21-year period. Table 1 shows the maximum cross-correlation values, $r$, and the time lags. The positive values mean the corresponding $a$-coefficients predominantly vary in phase with the solar cycle, while the negative values mean that the coefficients vary in anti-phase.  The uncertainties are calculated by performing Monte-Carlo simulations using the standard deviations of the averaged $a$-coefficients and the sunspot number.

\begin{table}[]
	\caption{ Statistics of cross correlation of the f-mode frequency splitting coefficients with the solar activity (sunspot number) showing the maximum absolute cross-correlations, the corresponding time lag and their phasing during the studied period of 1996-2017. The uncertainties are estimated by performing Monte-Carlo simulations. Positive values mean the coefficients vary mostly in phase with the sunspot number (their magnitude increases when the sunspot number increases). The negative sign indicates the coefficients vary mostly in anti-phase with the sunspot number.}
\label{table1}
\begin{tabular}{crrr}
\\
Splitting    & Cross-Correlation  	    &  	Time lag	\\
Coefficient  & Coefficient  $r$   &                            & \\
\hline\\
$a_1$ &	 $0.81\pm 0.05$	 &  $0.006\pm 0.049$       \\
$a_2$ &	 $-0.87\pm 0.04$  &  $0.095\pm 0.169$     \\
$a_3$ &	 $-0.51\pm 0.05$  &  $1.577\pm 0.587$    \\
$a_4$ &	 $0.60\pm 0.04$  &  $1.587\pm 0.126$     \\
$a_5$ &	 $-0.80\pm 0.07$  &  $1.074\pm 0.218$     \\
$a_6$ &     $0.76\pm 0.04$  &  $-1.564\pm 0.073$     \\
radius & $-0.94\pm 0.02$  &  $0.044\pm 0.019$  	  \\

\end{tabular}	
\end{table}

The results show that the magnitude of coefficients $a_1$, $a_4$, and $a_6$ grows in phase with the sunspot number (SN) with a relatively short time lag. But, coefficients $a_2$ and $a_5$ are in anti-phase with the SN. The $a_3$-coefficient, corresponding to the first differential rotation term, has the lowest cross-correlation among the other coefficients, but as noted above, its behavior is quite different from the others because it does not show a decrease corresponding to the current solar cycle. The $a_6$-coefficient correlates with the SN, but unlike the other coefficients it leads the SN by $\sim 1.6$ years. At this point, we do not have an explanation to this phenomenon. Also, the time period that had been studied by helioseismology is too short for a discussion of any predictive capabilities based on these coefficients. The most recent measurements of $a_6$ made in March-June of 2017 show that it probably has reached a minimum in the current cycle (Fig~\ref{asphericity}c). If this is the case then the next solar minimum can be expected 1.6 years later that is by the end of 2018 or in the beginning of 2019.

\section{The seismic radius}

The long standing question whether the solar radius is constant or not still remains in debate. There have been numerous studies, which led in the past to conflicting results. The reasons are known: the solar limb is not sharp as for a stainless ball, but is progressive due to the gaseous composition of the envelope. Thus, measuring with a high accuracy the diameter of the Sun is a challenge at the cutting edge of modern techniques. \citet{Rozelot2011} and \citet{Rozelot2016} gave a summary of efforts made from ancient time down to present days.
Measurements made from ground-based observations suffer from the atmospheric turbulence from which it is still hard to extract the solar signal. From space, the observations could be more favorable. However, the effects of the space environment (UV radiation, South Atlantic anomaly, thermal cycling, etc) combined with temperature variations, contamination by particles or radiation on the detector lead to degradation of the instruments in orbit, which directly observe the Sun. All these effects require corrections which prevents measurements from being well determined, as in the case of ground-based observations. Helioseismology permits, in principle, to find an alternative way for measuring the solar radius variations.

The f-mode frequencies  are sensitive to the sharp density gradient in the near-surface layers. They can be used for measurements of what is called the ``seismic radius" of the Sun, to differentiated it from the conventional photospheric radius. The procedure to derive the solar seismic radius has been described by \cite{Schou1997} and explained in details in \citet{Dziembowski2001}. Let us recall that  the seismic radius variations are  obtained from:
\begin{equation}
\frac{\Delta R}{R}=-\frac{2}{3}\left<\frac{\delta \nu}{\nu}\right>_\ell,
\label{eqradius}
\end{equation}
where the relative variations of f-mode frequency $\nu$ are averaged over a range of angular degree $\ell$. 
It follows from the equation derived by \citet{Dziembowski2004} who established a relation between the relative
frequency variations $\delta\nu/\nu$ for f-mode frequencies and the
associated Lagrangian perturbation of the radius $\delta r/r$ of
subsurface layers located at radius $r$:
\begin{equation}
\frac{\delta\nu_\ell}{\nu_\ell}=-\frac{3\ell}{2\omega_\ell^2 I_\ell}\int
\frac{g}{r}\frac{\delta r}{r}dI_\ell \label{eq_radius}
\end{equation}
where $\ell$ is the degree of the f modes, $I_\ell$ is the mode moment of inertia, which is proportional to the mode energy; $\omega_\ell=2\pi\nu_\ell$, and  $g$ is the gravity acceleration.

Equation  (4) leads to  relation (\ref{eqradius}) used in the previous determinations of the seismic radius  \citep{Schou1997}, assuming that $\delta r/r$ is constant with depth. However, these variations are not constant, and the value of the seismic radius depends on the exact localization of the f-mode energy beneath  the solar surface, which depends on the mode angular degree, $\ell$. Therefore, the estimate given by Eq.~(3) depends on the range of angular degree of the averaged mode frequencies. In addition, the frequency variations induced by the surface magnetic activity affect inferences of the seismic radius variations \citep{Dziembowski2001}. 

The result obtained by averaging the f-mode frequency variations for the whole common subset of modes in the whole observed angular degree range, $\ell=137-299$, is shown in Fig.~\ref{radius}$b$. It represents a proxy of the seismic radius relative variations with the modulation amplitude of about $-2.3\times 10^{-5}$ in Solar Cycle 23 and $-1.7\times 10^{-5}$ in Cycle 24. It is clear that the seismic radius varies in the  phase opposite to the solar cycle with a very small time lag (Table~1). When the averaging includes the f-modes of $\ell=137-200$ the seismic radius modulations are as twice as small, but the phase relation remains. 

\citet{Lefebvre2005} and \citet{Lefebvre2007} argued that the radius variations are not constant and non-homologous in the subsurface layers, and attempted to reconstruct the depth dependence of $\delta r/r$ with depth. Such radius variations can give a real insight into changes of the Sun's subsurface stratification. This analysis needs to be repeated for the new data.

\section{Conclusion}

The helioseismology data from the SoHO and SDO space missions have continuously covered the whole Solar Cycle 23 and most of Solar Cycle 24, providing a unique opportunity to perform comparable studies of the two solar cycles that are significantly different in the level of magnetic activity, and investigate solar-cycle variations of the subsurface structure and dynamics. In this article we presented an analysis of properties of the surface gravity modes of solar oscillations (called f-mode). These modes in the medium-degree range are trapped in the subsurface layer $~\sim 15$~Mm deep. This layer is located in the near-surface rotational shear layer (NSSL) which is believed to play a key role in the solar-cycle dynamo mechanism. The results showed that the f-mode splitting coefficients exhibit clear solar-cycle variations, providing information about changes in the subsurface differential rotation and the solar asphericity. A particularly unexpected feature is the absence of a decrease of the $a_3$ splitting coefficient, which describes the main differential rotation term,  during Cycle 24. This potentially reflects a long-term trend of the future solar cycles. It is clear that such data  provide a considerable help to understand the underlying mechanisms of the solar cycle variations and physical processes in the Near Sub-Surface Layers of the Sun.\\

\noindent
{\bf Acknowledgments.} The work was performed with the support of the International Space Science Institute (ISSI) in Bern (CH), the VarSITI (Variability of the Sun and Its Terrestrial Impact) Program of the Scientific Committee On Solar-Terrestrial Physics (SCOSTEP). The authors thank the ISSI for holding a scientific meeting on solar variability organized by K. Georgieva. The work was partially supported by NASA grants  NNX14AB70G and\\ NNX17AE76A.\\


\noindent
{\bf References}


\end{document}